\documentclass[11pt]{article}
\usepackage{amsmath}
\usepackage{amsfonts}
\usepackage{amssymb}
\usepackage{amsmath}
\usepackage{amsfonts}
\usepackage{amssymb}
\usepackage{wrapfig}
\usepackage{graphicx}
\begin{document}
\begin{center}
\begin{Large} {\bf Some Theoretical Aspects of Observation of Acceleration Induced Thermality}
\end{Large}
\vspace{10mm} \\
\begin{Large}
{Yefim S. Levin }\\
Department of  Mathematics, Salem State University,
Salem MA, 01970   \\
\end{Large}
\end{center} 
\begin{abstract}
In recent work by M.H.Lynch, E.Cohen, Y.Hadad and I.Kaminer (LCHK), a modified model of the Unruh-DeWitt two-level quantum detector, coupled to a semi-classical 4-vector current, has been proposed  to examine the radiation emitted by high energy positrons channeled into silicon crystal samples. Inspired by their ideas, we analyze theoretical aspects of such a model, its internal consistency,
and ignore all questions related to experiments.
 The two-potential correlation functions for the quantized electromagnetic field in a vacuum state and the corresponding detector radiation power (DRP), considered in proper time formalism, are used as the basis for investigating the radiation observed at an accelerating point detector. 
  The quantum detector is assumed to be moving through an electromagnetic vacuum along a classical hyperbolic trajectory with a constant proper acceleration. The DRP is obtained for three possible cases. First, the DRP is found in a Lorentz-invariant manner. It contains both transverse and non-physical longitudinal polarization modes and is a divergent quantity. 
 Second, the radiation power holds only physical transverse modes but it is non-relativistic and also depends on the detector proper time, which contradicts the fact that there is no preferred time for hyperbolic detector motion. Third, in the particular case considered by LCHK, for zero detector proper time when its velocity in the lab inertial system is zero, the radiation power with transverse modes shows some signs of thermality which could be associated with a detector acceleration but different from the Bose-Einstein statistics expected for the photon field. If the detector energy gap is zero then, in complete contradiction with what LCHK claim, there is no radiation and no "thermalized Larmor formula". Based on our analysis we do not believe that the LCHK's model 
 can be used to support the idea about thermal effects of uniform acceleration.
{\large }\end{abstract} 
\section{ Introduction }
\indent The theoretical prediction of a warmed vacuum observed by an accelerated observer (\cite{1975_Davis},\cite{1976_Unruh},\cite{1982_Birrell}) was a by-product of the extension of quantum theory to curved spaces. P.W. Milonni found this result so remarkable that much later, in 1994, he wrote in his book (\cite{1994_Milonni}p 64): "It took half a century after the birth of the quantum theory of radiation for the thermal effect of uniform acceleration to be discovered." He provides an elegant description of the Unruh-Davis effect in a scalar field done in terms of the correlation function. The concept of an accelerating observer is given there in general terms, specifying only its classical trajectory and without indicating its internal structure and its interaction with the vacuum. However, in any experimental situation you cannot avoid this issue. This situation is clearly formulated by B.S. DeWitt (\cite{1979_DeWitt}): "These and many examples discovered in the last few years have brought about major changes in our ways of thinking about particles and in our ways of defining them and their associated vacuum states. As has happened before in relativity theory, and in the quantum theory, one has had to fall back on \emph{operational definitions} \footnote{Emphasized by the author}. Here, for example, one must ask: How would a given particle detector respond in a given situation." Today, when there is a great interest in an experimental proof of the effect, this remark becomes relevant again. What theoretical implementation and what kind of a detector model can be used to explain experiments? \\
\indent Recently, Cozzella et al.(\cite{2017_Cozzella}) made a proposal for observing the Unruh-Davis effect, which outlined a method of measuring the Fulling-Davis-Unruh (FDU) temperature directly from a data set. There is an opinion (\cite{2021_Lynch},\cite{2022_Lynch}) that experimental evidence already exists. This opinion is based on both experimental data and a theoretical model for their interpretation.
 An essential part of their analysis is a slightly changed Unruh-DeWitt detector model.
The latter was proposed for an observation of  a scalar field.\\
\indent For the case of an electromagnetic field, LCHK use the Unruh-DeWitt detector, coupled to the semi-classical vector current:
\begin{eqnarray}
\label{eq:current}
\hat{j}_{\mu}=u_{\mu}\hat{q}(\tau)\delta^3(x-x_{tr})
\end{eqnarray}
and the current interaction for QED
\begin{eqnarray}
\hat{S}_{I}=\int d^4 x \hat{j}_{\mu}(x)\hat{A}^{\mu}(x),
\end{eqnarray}
where $u_{\mu}$ is the 4-velocity of the detector, the monopole operator $\hat{q}(\tau)$ is Heisenberg evolved via $\hat{q}(\tau)=e^{i\hat{H} \tau}\hat{q}(0)e^{-i\hat{H} \tau}$, 
with $\hat{q}(0)$ defined as $\hat{q}(0) |E_i>=|E_f> $ and with $E_i$ and $E_f$ the initial and final energy of a two energy state detector moving along the trajectory, $x_{tr}(\tau)$.\\
\indent  Inspired by LCHK's ideas, we analyze theoretical aspects of such a model, its internal consistency, and ignore all questions related to experiments. For this purpose we slightly modify their model and explicitly use proper time formalism. Our model is described in details in Section \ref{Detector Model}.\\
\indent The main tool of our investigation is a calculation of a two-potential
correlation function of the potentials at the locations of the detector on a hyperbolic trajectory at two proper times, $\tau_1$ and $\tau_2$, for the vacuum state of an electromagnetic field determined at the lab inertial system in the Minkowski space-time. This is done in a relativistic invariant manner in Subsection \ref{The Correlation Function in Terms of }. Calculated this way, the correlation function
can be split, as shown in Subsection \ref{The Correlation Function:Scalar/Longitudinal and Transverse Polarization Modes}, into two parts corresponding respectively to transverse polarization and scalar/longitudinal modes. In Subsection \ref{supplementary
 condition}, the transverse part is obtained by explicitly using the supplementary Lorentz condition. \\
\indent To be an observable in a real experiment, the detector radiation power (DRP) is a more convenient variable than just a correlation function. In Section \ref{Detector Radiation Power.} we consider 3 cases. First, relativistic DRP is discussed in Subsection \ref{relativistic DRP}. Second, DPR consisting of transverse polarization modes is obtained and discussed in Subsection \ref{transverse DRP}. Third, derivation and discussion of LCHK's DRP can be found in Subsection \ref{The Thermal Larmor Formula}. \\
\indent The author intends to focus only on theoretical validation of LCHK' model, leaving aside other work in this area. As a result, the bibliography contains only references which can be associated with LCHK's work.               
\section{Detector Model}
\label{Detector Model}
\indent A point-like quantum detector, which can be in one of two quantum states, is uniformly accelerated along a classical hyperbolic trajectory in a quantized electromagnetic field which is in a vacuum state. Our goal is to find the probability of detector transition from one of its states to the other and its radiation power, induced by the interaction of the detector with the vacuum and  accompanied by all possible one-photon states of radiation. \\
\indent In our analysis we used a fixed inertial frame $I_M$ in Minkowski 4-space and a set of inertial frames $I_{\tau}$ \cite{1980_Boyer},\cite{1985_Cole} defined by the condition that the detector is instantaneously at rest at $I_{\tau}$ at the proper time $\tau$ measured by a clock at the detector position.
In each $I_{\tau}$-frame the detector labeled by $(t_{\tau},x_{\tau},y_{\tau}, z_{\tau})=(\tau,x_{\tau},0,0)$ has the same  acceleration, $d^2x_{\tau}/dt^2_{\tau}=a=const$.\\
\indent Its trajectory in the Minkowski space is \footnote{Constants c and $\hbar$ are set to 1}
\begin{eqnarray}
\label{eq:Trajectory}
x(t)=(\frac{1}{a})(1 +a^2 t^2)^{1/2}
\end{eqnarray}
or in terms of proper time $\tau$    
\begin{eqnarray}
\label{eq:trajectory}
t(\tau)=\frac{1}{a} \sinh(a\tau),\;\; x(\tau)=\frac{1}{a} \cosh(a\tau),\;\; y(\tau)=z(\tau)=0,
\nonumber \\
v(\tau)= \tanh(a\tau),\;\; d^2 x /d t^2= \gamma_{\tau}^{-3}a,
\end{eqnarray} 
where $\gamma_{\tau}$ is the Lorentz boost of a local Lorentz transformation between $I_M$ and $I_{\tau}$
\begin{eqnarray}
\label{eq:boost}
\gamma(\tau)=\frac{1}{(1 -v^2(\tau))^{1/2}}=\cosh(a\tau),\;\; 
\end{eqnarray}
The Unruh-DeWitt detector is coupled to the vector current operator, according to our definition, in $I_{\tau}$ frame \footnote{In \cite{2021_Lynch} the current is defined in $I_M$ frame with the help of the $\delta$-function: $\hat{j}_{\mu}=u_{\mu}\hat{q}(\tau)\delta^3(x-x_{tr(\tau)})$. }:
\begin{eqnarray}
\hat{j}_{\tau,\mu}(\tau,x_{\tau},0,0)= u_{\tau,\mu}(\tau,0,0,0) \hat{q}(\tau), \;\; \mu=0,1,2,3, \nonumber \\
u_{\tau,\mu}\equiv (u_{\tau 0},u_{\tau 1},u_{\tau 2},u_{\tau 3})=(1,0,0,0)
\end{eqnarray}
Here $u_{\tau,\mu}$ is 4-velocity of the detector in $I_{\tau}.$  \\
\indent The Hamiltonian of the detector's interaction with an electromagnetic field in $I_{\tau}$ is 
\begin{eqnarray}
\hat{H}_i(\tau)= \hat{j}_{\tau,\mu}(\tau) \hat{A}_{\tau,\mu}(\tau,0)=\hat{j}_{\tau,0}(\tau) \hat{A}_{\tau,0}(\tau,0)=\hat{q}(\tau)\hat{A}_{\tau,0}(\tau,0).
\end{eqnarray}
It depends on the zeroth component of the 4-potential, $\hat{A}_{\tau,0}(\tau,0)$, defined at the location of the detector in $I_{\tau}$ at the proper time $\tau$.  The amplitude of the transition from $|E_i,0 >$ to $|E_f, \vec{k}\lambda>$ in the first order of perturbation is \footnote{A proper time formalism is used here as in \cite{1982_Birrell}, Section 3.3.}
\begin{eqnarray}
A_{E_f,\vec{k},\lambda; E_i,0 }
	=-i < E_f, \vec{k}\lambda |\;\int_{-\infty}^{\infty} d \tau \;\hat{q}(\tau) \hat{A}_{\tau,0}(\tau,0)\; |E_i,0> = \\
=-i <E_f|\;\hat{q}(0)\;|E_i> \int_{-\infty}^{\infty} d\tau\; e^{i(E_f-E_i) \tau}<\vec{k}\lambda |\; \hat{A}_{\tau,0}(\tau,0)\; |0>.
\end{eqnarray}
Here $|E_i, 0>$ is a detector state with energy $E_i$, the electromagnetic field in a vacuum state, and $|E_f, \vec{k} \lambda>$ is the state with energy $E_f$ and the electromagnetic field in a one-photon state with momentum $\vec{k}$ and polarization mode $\lambda$. The states $|E_i>$,$|E_F>$ and the states $|\vec{k}, \lambda>$, $|0>$ are defined in different reference
systems, $I_{\tau}$ and $I_M$ respectively. \\  
\indent The total probability of the detector transition from the state with energy $E_i$ to the state with energy $E_f$, after summation over final one-photon states with all possible $\vec{k}$ and polarizations $\lambda$, is 
\begin{eqnarray}
|A|^2 \equiv q^2\;\int d\tau_1 d\tau_2 \exp{[-i\Delta E\;( \tau_2-\tau_1)]}\int d\vec{k}\;\sum_{\lambda=0}^{\lambda=3}\;<0|(\hat{A}_{\tau,0})^{\dagger}(\tau_2,0)|\vec{k}\lambda><\vec{k}\lambda|\hat{A}_{\tau,0}(\tau_1,0)|0> \equiv \nonumber \\
\equiv q^2\;\int d\tau_1 d\tau_2 \exp{[-i\Delta E\; (\tau_2-\tau_1)]} <0|\; (\hat{A}_{\tau,0})^{\dagger}(\tau_2,0)\hat{A}_{\tau,0}(\tau_1,0)|0>  \nonumber \\
\end{eqnarray}
or \footnote{The $\tau$ index should not be confused with the integration variable, $\tau$. } in new variables
\begin{eqnarray}
\label{eq:total probability}	
|A|^2= q^2\;\int d\sigma d\tau \exp{[-i\Delta E\; \tau]}\times 
 <0|\; (\hat{A}_{\tau,0})^{\dagger}(\sigma + \tau/2,0)\hat{A}_{\tau,0}(\sigma-\tau/2,0)|0>,
\end{eqnarray}
where
\begin{eqnarray}
q^2=|<E_{f}| \hat{q}(0)|E_i>|^2, \,\,\,  \Delta E=E_f-E_i, \;\;\; \sigma =(\tau_2 +\tau_1)/2,\;\;\; \tau=\tau_2-\tau_1
\end{eqnarray}
\indent Summation over $\lambda$ in the previous expression should include both transverse modes, with $\lambda=1,2$, and scalar/longitudinal ones, with $\lambda=0,3$. Separation of them is possible in a fixed reference frame only but in proper time formalism a reference frame can not be fixed. We will see in the next section that calculation of the correlation function involves three reference systems, $I_{\tau_1}$, $I_{\tau_2}$ and $I_M$.
\section{The Correlation Function}
\label{CorFunct}
\subsection{The Correlation Function in Terms of Variables Determined in the Minkowski Space Time.}
\label{The Correlation Function in Terms of }
\indent The vacuum state, $|0 >$, and one-photon states, $|\vec{k}\lambda>$, are defined in $I_M$ but $\hat{A}_{\tau,0}(\sigma +\tau/2,0)$ and $\hat{A}_{\tau,0}(\sigma -\tau/2,0)$ are defined at detector locations in two different instantaneous  inertial reference frames. Therefore we have to represent $\hat{A}_{\tau, 0}(\sigma\pm\tau/2,0)$, the scalar component of 4-potential, in terms of 4-potential operator components in $I_M$ using Lorentz transformations with boosts $\gamma_{\sigma+ \tau/2}$ and $\gamma_{\sigma- \tau/2}$ ( \ref{eq:boost}) respectively:
\begin{eqnarray}
\label{eq:potential transform}
\hat{A}_{\tau,0}(\sigma \pm  \tau/2, 0)=\hat{A}_0[t(\sigma \pm \tau/2),x(\sigma \pm \tau/2), 0,0]\gamma_{\sigma \pm \tau/2}-
\hat{A}_1[t(\sigma \pm \tau/2),x(\sigma \pm \tau/2), 0,0](v_{\sigma \pm \tau/2}/c)\gamma_{\sigma \pm \tau/2}= \nonumber \\
\hat{A}_0[t(\sigma \pm \tau/2),x(\sigma \pm \tau/2), 0,0] \cosh[a/c(\sigma\pm \tau/2)]-
\hat{A}_1[t(\sigma \pm \tau/2),x(\sigma \pm \tau/2), 0,0] \sinh[a/c(\sigma\pm\tau/2)] \nonumber \\
\end{eqnarray}
\indent
Using the expansion of $A_{\mu}(t,x,0,0)$ in plain waves in $I_M$ (\ref{eq:expansion}), properties of annihilation/creation operators (\ref{eq:cr_an_operators}) and trajectory equations (\ref{eq:trajectory})
this correlation function can be reduced to the form:
\begin{eqnarray}
\label{eq:corfunction}
<0|\; \hat{A}_{\tau,0})^{\dagger}(\sigma +\tau/2,0)\hat{A}_{\tau,0}(\sigma -\tau/2,0)|0>
= \nonumber\\
-\int \frac{d^3k}{(2 \pi)^3}\frac{1}{2 \omega_{k}}
\exp\{-i [\omega_k (t_{\sigma +\tau/2}-t_{\sigma -\tau/2}) - k_x (x_{\sigma +\tau/2}-x_{\sigma -\tau/2})]\}
 \times \nonumber \\ 
\nonumber \\
\times \{
+\cosh(a(\sigma+\tau/2))\cosh(a(\sigma-\tau/2)) \sum_{\lambda=0}^3 [e_0(\vec{k},\lambda)]^2g^{\lambda \lambda}+ \nonumber \\
-\cosh(a(\sigma+\tau/2))\sinh(a(\sigma-\tau/2)) \sum_{\lambda=0}^3 [e_0(\vec{k},\lambda)][e_1(\vec{k},\lambda)] g^{\lambda \lambda} -\nonumber \\
-\sinh(a(\sigma+\tau/2))\cosh(a(\sigma-\tau/2)) \sum_{\lambda=0}^3 [e_1(\vec{k},\lambda)][e_0(\vec{k},\lambda)] g^{\lambda \lambda}+ \nonumber \\
+\sinh(a(\sigma+\tau/2))\sinh(a(\sigma-\tau/2)) \sum_{\lambda=0}^3 [e_1(\vec{k},\lambda)]^2 g^{\lambda \lambda} \nonumber \\ 
\} 
\end{eqnarray} 
For our further analysis, it is useful to consider the contribution to the correlation function from scalar/longitudinal and transverse polarization modes of the detector radiation. It is done in the next subsection.
\subsection{Scalar/Longitudinal and Transverse Polarization Modes}
\label{The Correlation Function:Scalar/Longitudinal and Transverse Polarization Modes}
With the help of the properties of polarization vectors (\ref{eq:polarization vectors}) and hyperbolic function  (\ref{eq:hyperbolic functions}), provided in  Appendix \ref{Vector Potentials}, the correlation function can be split into two parts corresponding to scalar/longitudinal (sl) and transverse (tr) polarization modes
\begin{eqnarray}
\label{eq:CF}
<0|\; \hat{A}_{\tau,0})^{\dagger}(\sigma +\tau/2,0)\hat{A}_{\tau,0}(\sigma -\tau/2,0)|0>
= \nonumber\\
-\int \frac{d^3k}{(2 \pi)^3}\frac{1}{2 \omega_{k}}
\exp\{-i [\omega_k (t_{\sigma +\tau/2}-t_{\sigma -\tau/2}) - k_x (x_{\sigma +\tau/2}-x_{\sigma -\tau/2})]\} 
\times \{()_{sl} + ()_{tr}\},
\end{eqnarray}
where
\begin{eqnarray}
\label{eq:CFsl}
()_{sl}\equiv
\biggl[ 
\frac{1}{2}\cosh(2a\sigma) + \frac{1}{2}\cosh(2a\tau)
\biggr][e_0(\vec{k},\lambda=0)]^2-
\biggl[ 
\frac{1}{2}\cosh(2a\sigma) - \frac{1}{2}\cosh(2a\tau)
\biggr][e_1(\vec{k},\lambda=3)]^2= \nonumber \\
=\frac{1}{2}\cosh(2a\sigma)\sin^2\theta+\frac{1}{2}\cosh(a\tau)(1+ \cos^2\theta), \nonumber \\
\end{eqnarray}
\begin{eqnarray}
\label{eq:CFtr}
()_{tr}\equiv
\biggl[ 
\frac{1}{2}\cosh(2a\sigma) - \frac{1}{2}\cosh(2a\tau)
\biggr]([e_1(\vec{k},\lambda=1)]^2+[e_1(\vec{k},\lambda=2)]^2) = \nonumber \\
=-\frac{1}{2}\cosh(2a\sigma)\sin^2\theta+\frac{1}{2}\cosh(a\tau)\sin^2\theta
\nonumber \\
\end{eqnarray}
and $\theta$ is an angle between the wave vector $\vec{k}$ and the detector motion direction along axis x. \\ 
\indent Obviously the splitting of the correlation function into sl- and tr-parts is not Lorentz invariant and depends on $\sigma$.
The $\sigma$-term in the correlation function contains contributions from both scalar/longitudinal and transverse modes, which are mutually compensated in the sum of $()_{sl}$ and $()_{tr}$
\begin{eqnarray}
\label{eq:slplustr}
()_{sl} + ()_{tr}= \cosh(a \tau)
\end{eqnarray}
 and
\begin{eqnarray}
<0|\; \hat{A}_{\tau,0})^{\dagger}(\sigma +\tau/2,0)\hat{A}_{\tau,0}(\sigma -\tau/2,0)|0>
= \nonumber\\
-\int \frac{d^3k}{(2 \pi)^3}\frac{1}{2 \omega_{k}}
\exp\{-i [\omega_k (t_{\sigma +\tau/2}-t_{\sigma -\tau/2}) - k_x (x_{\sigma +\tau/2}-x_{\sigma -\tau/2})]\} 
\times \cosh(a \tau)
\nonumber \\ 
\end{eqnarray}
\indent The factor $\cosh(\frac{a}{c})\tau$ does not depend on $\sigma$. 
We should expect that the correlation function does not depend on $\sigma$ as well because there is no preferred time for hyperbolic motion. The independence on $\sigma$ can be exhibited explicitly by changing the variables of integration $\vec{k}$ to $\vec{k}^{\prime}$, where 
\begin{eqnarray}
\label{eq:variables_change}
\omega=\omega^{\prime} \cosh(a\sigma)+k_x^{\prime}\sinh(a\sigma), \nonumber \\
k_x=k_x^{\prime}\cosh(a\sigma)+\omega^{\prime}\sinh(a\sigma), \nonumber \\
k_y=k_y^{\prime}, \;\;\; k_z=k_z^{\prime}, \;\;\; \int \frac{d\vec{k}}{(2 \pi)^3} \frac{1}{\omega}= \int\frac{d\vec{k^{\prime}}}{(2 \pi)^3} \frac{1}{\omega^{\prime}},
\end{eqnarray}  
which corresponds exactly to a Lorentz transformation from the unprimed laboratory frame $I_M$ over to the primed inertial frame 
in which the accelerating detector is instantaneously at rest at proper time $\sigma$.
Then we obtain
\begin{eqnarray}
\label{eq:correlation function integral }
<0|\; \hat{A}_{\tau,0})^{\dagger}(\sigma+\tau/2,0)\hat{A}_{\tau,0}(\sigma-\tau/2,0)|0>\equiv 
<0|\; \hat{A}_{\tau,0})^{\dagger}(\tau/2,0)\hat{A}_{\tau,0}(\tau/2,0)|0> = \nonumber \\
=-\int \frac{d^3k^{\prime}}{(2 \pi)^3}\frac{1}{2 \omega^{\prime}} \exp\{-i \frac{2}{a}\omega^{\prime}\sinh(\frac{a}{2}\tau)\} \cosh(a\tau).
\end{eqnarray}
This result is obtained in a relativistically invariant manner but it contains  nonphysical  , scalar/longitudinal, polarization modes. Now we will find the correlation function without unphysical modes.  
\subsection{Supplementary Condition}
\label{supplementary condition}
Until now we considered all polarization modes of the quantized electromagnetic field on an equal basis even though nonphysical scalar and longitudinal modes should be excluded in the quantized field. Now we explicitly introduce the supplementary condition to exclude nonphysical  modes. For this purpose we will use the gauge transformation \cite{1961_Schweber}(9.41) in Gupta-Bleyer formalism  
\begin{eqnarray}
\label{eq:gauge transformation}
a^{\dagger}_{\mu}(\vec{k}) \rightarrow \tilde{a}^{\dagger}_{\mu}(\vec{k})=a^{\dagger}_{\mu}(\vec{k})-\frac{k_{\mu}}{k_0}a^{\dagger}_0(\vec{k}), \nonumber \\
a_{\mu}(\vec{k}) \rightarrow \tilde{a}_{\mu}(\vec{k})=a_{\mu}(\vec{k})-\frac{k_{\mu}}{k_0}a_0(\vec{k}) \\
a_{\mu}(\vec{k})\equiv \sum_{\lambda=0}^{3}a(\vec{k},\lambda)e_{\mu}(\vec{k},\lambda),\;\;\;
a_{\mu}^{\dagger}(\vec{k})\equiv \sum_{\lambda=0}^{3}a^{\dagger}(\vec{k},\lambda)e_{\mu}(\vec{k},\lambda)
\end{eqnarray}  
of the 4-vector potential operator defined in $I_M$
\begin{eqnarray}
\hat{A}_{\mu}(t,\vec{x})=\int \frac{d^3k}{(2 \pi)^3}\frac{1}{2 \omega_k}[a_{\mu}(\vec{k})e^{-kx} +a_{\mu}^{\dagger}(\vec{k})e^{kx}]
\end{eqnarray} 
to
\begin{eqnarray}
\label{eq:after gauge transformation}
\tilde{\hat{A}}_{\mu}(t,\vec{x})=\int \frac{d^3k}{(2 \pi)^3}\frac{1}{2 \omega_k}[\tilde{a}_{\mu}(\vec{k})e^{-kx} +\tilde{a}^{\dagger}_{\mu}(\vec{k})e^{kx}].
\end{eqnarray}
Then, after this gauge transformation performed in $I_M$,  instead of (\ref{eq:potential transform}) and (\ref{eq:corfunction}) we have respectively
\begin{eqnarray}
\label{eq:potential transform 1}
\hat{A}_{\tau,0}(\sigma \pm  \tau/2, 0)=\tilde{\hat{A}}_0[t(\sigma \pm \tau/2),x(\sigma \pm \tau/2), 0,0]\gamma_{\sigma \pm \tau/2}-
\tilde{\hat{A}}_1[t(\sigma \pm \tau/2),x(\sigma \pm \tau/2), 0,0](v_{\sigma \pm \tau/2}/c)\gamma_{\sigma \pm \tau/2}= \nonumber \\
\tilde{\hat{A}}_0[t(\sigma \pm \tau/2),x(\sigma \pm \tau/2), 0,0] \cosh[a/c(\sigma\pm \tau/2)]-
\tilde{\hat{A}}_1[t(\sigma \pm \tau/2),x(\sigma \pm \tau/2), 0,0] \sinh[a/c(\sigma\pm\tau/2)] \nonumber \\
\end{eqnarray}
and 
\begin{eqnarray}
<0|\; \hat{A}_{\tau,0})^{\dagger}(\sigma +\tau/2,0)\hat{A}_{\tau,0}(\sigma -\tau/2,0)|0>
= \nonumber\\
\int \frac{d^3k}{(2 \pi)^3}\frac{1}{2 k_0}
\times \frac{d^3 k^{\prime}}{(2 \pi)^3}\frac{1}{2 k_0^{\prime}} \times \nonumber \\
\times \exp\{-i [k_0^{\prime} t_{\sigma +\tau/2} - k_1^{\prime} x_{\sigma +\tau/2}]+ i[k_0 t_{\sigma -\tau/2} - k_1 x_{\sigma -\tau/2}]
\}
\times \nonumber \\ 
\times \{
+\cosh(a(\sigma+\tau/2))\cosh(a(\sigma-\tau/2))<0|\tilde{a}_0(\vec{k}^{\prime}) \tilde{a}^{\dagger}_0(\vec{k})|0> - \nonumber \\
-\cosh(a(\sigma+\tau/2))\sinh(a(\sigma-\tau/2))<0|\tilde{a}_0(\vec{k}^{\prime}) \tilde{a}^{\dagger}_1(\vec{k})|0> -\nonumber \\
-\sinh(a(\sigma+\tau/2))\cosh(a(\sigma-\tau/2))<0|\tilde{a}_1(\vec{k}^{\prime}) \tilde{a}^{\dagger}_0(\vec{k})|0> + \nonumber \\
+\sinh(a(\sigma+\tau/2))\sinh(a(\sigma-\tau/1))<0|\tilde{a}_1(\vec{k}^{\prime}) \tilde{a}^{\dagger}_1(\vec{k})|0>. \nonumber \\ 
\} 
\end{eqnarray} 
Taking into account (\ref{eq:gauge transformation}) and (\ref{eq:cr_an_operators}) we have
\begin{eqnarray}
[a_{\mu}(\vec{k}), a^{\dagger}_{\nu}(\vec{k}^{\prime})]=-(2\pi)^3 2 k_0\; g_{\mu \nu} \;\delta^3(\vec{k}-\vec{k}^{\prime})
\end{eqnarray}
and
\begin{eqnarray}
<0|\tilde{a}_0(\vec{k}^{\prime}) \tilde{a}^{\dagger}_0(\vec{k})|0> =<0|\tilde{a}_0(\vec{k}^{\prime}) \tilde{a}^{\dagger}_1(\vec{k})|0>=
<0|\tilde{a}_1(\vec{k}^{\prime}) \tilde{a}^{\dagger}_0(\vec{k})|0> =0,\nonumber \\
<0|\tilde{a}_1(\vec{k}^{\prime}) \tilde{a}^{\dagger}_1(\vec{k})|0> =
(2\pi)^3\;2k_0 (1-\frac{k_1^2}{k_0^2}) \delta^3(\vec{k}-\vec{k}^{\prime})
\end{eqnarray}
Then the correlation function takes the form
\begin{eqnarray}
<0|\; \hat{A}_{\tau,0})^{\dagger}(\sigma +\tau/2,0)\hat{A}_{\tau,0}(\sigma -\tau/2,0)|0>
= \nonumber\\
-\frac{1}{(2 \pi)^3} \frac{1}{4} [2 \cosh^2(a\sigma)-1-\cosh(a\tau)]
\int \frac{d^3k}{(2 \pi)^3}\frac{1}{k_0} (1-\frac{k_1^2}{k_0^2}) \times \nonumber \\
\times\exp\{-i [k_0 (t_{\sigma +\tau/2}-t_{\sigma -\tau/2}) - k_1 (x_{\sigma +\tau/2}-x_{\sigma -\tau/2})]\}
\end{eqnarray} 
or
\begin{eqnarray}
<0|\; \hat{A}_{\tau,0})^{\dagger}(\sigma +\tau/2,0) \hat{A}_{\tau,0}(\sigma -\tau/2,0)|0>
= \nonumber\\
\frac{1}{4} [2 \cosh^2(a\sigma)-1-\cosh(a\tau)]
\int \frac{d^3k}{(2 \pi)^3}\frac{1}{k_0}\sin^2 \theta
  \times \nonumber \\
\times\exp\{-i [k_0 (t_{\sigma +\tau/2}-t_{\sigma -\tau/2}) - k_1 (x_{\sigma +\tau/2}-x_{\sigma -\tau/2})]\},
\end{eqnarray} 
where $\theta$ is an angle between $\vec{k}$ and the direction of the detector motion. \\
\indent It is obvious that the correlation function is exactly the transverse part of the correlation function obtained in  (\ref{eq:CF}) and (\ref{eq:CFtr}), as it is supposed to be.
\section{Detector Radition Power}
\label{Detector Radiation Power.}
Detector radiation power for correlation function (\ref{eq:correlation function integral }) contains nonphysical polarization modes
(that do not correspond to a real electromagnetic field) but it is relativistic and will be useful for interpretation of the results obtained under the supplementary condition when unphysical modes are annihilated. Therefore we  consider both cases: for correlation functions  (\ref{eq:correlation function integral }) with unphysical states, and the function defined by (\ref{eq:CF}) and (\ref{eq:CFtr}) without unphysical states. Then we discuss the approach adapted by LCHK to arrive at their expression for DRP.  
\subsection{Relativistic DRP }
\label{relativistic DRP}
\indent From equation (\ref{eq:correlation function integral }) and (\ref{eq:total probability}) we have 
 \begin{eqnarray}
 \frac{d |A|^2}{d \sigma} =-q^2\int_{-\infty}^{+\infty}d\tau \exp(-i \Delta E \tau)\int \frac{d^3k^{\prime}}{(2 \pi)^3}\frac{1}{2 \omega^{\prime}}\times
  \exp\{-i \frac{2}{a}\omega^{\prime}\sinh(\frac{a}{2}\tau)\}\times \cosh(a\tau)
 \end{eqnarray}
This is the probability of the detector  transition from $|E_i>$ to $|E_f>$ per unit proper time, $\sigma$, accompanied by one-photon radiation of all possible  $\vec{k}^{\prime}$ and polarizations $\lambda=0,1,2,3$. The corresponding electromagnetic energy radiation, $W,$ per unit proper time, involves an additional factor $\omega^{\prime}$ in the last integrand 
\begin{eqnarray}
\label{eq:power}
S\equiv \frac{dW}{d\sigma}= -q^2\int_{-\infty}^{+\infty}d\tau \exp(-i \Delta E \tau)\int \frac{d^3k^{\prime}}{(2 \pi)^3}\frac{1}{2 \omega^{\prime}}\times
\omega^{\prime} \exp\{-i \frac{2}{a}\omega^{\prime}\sinh(\frac{a}{2}\tau)\}\times \cosh(a\tau)
 \end{eqnarray}
and after integration over $\vec{k}^{\prime}$, it is
\begin{eqnarray}
\label{eq:detector radiation power }
S= -\frac{i}{4 \pi}\alpha a^3 \int_{-\infty}^{+\infty}d \tau \frac{e^{-i \Delta E \tau} \cosh(a\tau)}{\sinh^3(\frac{a \tau}{2})} \equiv -\frac{i}{4 \pi} \alpha a^3 (I_1+I_2),
\;\;\; \alpha=\frac{q^2}{4 \pi}, 
\end{eqnarray}
where
\begin{eqnarray} 
I_1 \equiv \int_{-\infty}^{+\infty}d \tau \frac{\cos (\Delta E \tau ) \cosh(a\tau)}{\sinh^3(\frac{a \tau}{2})}, \;\;\;
\label{divergent integral}
I_2 \equiv \int_{-\infty}^{+\infty}d \tau \frac{-i \sin( \Delta E \tau) \cosh(a\tau)}{\sinh^3(\frac{a \tau}{2})}
\end{eqnarray}
This expression is Lorentz invariant and does not depend on $\sigma$. It means that any observer which is at rest at the location of the detector measures the same radiation power. This good feature of the model is obscured by two facts. \\
\indent The first is that S is divergent. Indeed, the integral $I_1=0$ because its integrand is an odd function but $I_2$ is divergent (Appendix \ref{Divergence of I2}). \\
\indent The second fact is that we have not taken into consideration the supplementary condition to get rid of unphysical polarization modes.
This will be done in the next section.   
\subsection{Transverse DRP}
\label{transverse DRP}
In this section we consider the correlation function obtained with the use of the supplementary condition. It contains only the transverse polarization modes.  \\
\indent From ( \ref{eq:total probability})  and (\ref{eq:CF}), (\ref{eq:CFtr}) we have 
\begin{eqnarray}
\frac{d |A|^2_{tr}}{d \sigma}=q^2 \frac{1}{4 (2 \pi)^3} \int d \tau  e^{-i \Delta E \tau} 
[ \cosh (2a \sigma) - \cosh (a \tau)]
\int d^3 k \frac{1}{\omega_k} \sin^2 \theta \times \nonumber \\
\times exp \{-i [\omega_k(t_{\sigma + \tau/2}- t_{\sigma - \tau/2}) -k_x(x_{\sigma + \tau/2}- x_{\sigma - \tau/2})\;] \;\}\\
\omega_k= k_0 \nonumber
\end{eqnarray}
and, after multiplication of the second integrand by $\omega_k$,  the detector radiation power due to transverse polarization modes is 
\begin{eqnarray}
\label{eq:transverse DRP}
S_{tr}(\sigma) \equiv \frac{d W_{tr}}{d \sigma}(\sigma) =q^2 \frac{1}{4 (2 \pi)^3} \int d \tau  e^{-i \Delta E \tau} 
[ \cosh (2a \sigma) - \cosh (a \tau)]
\int d^3 k  \sin^2 \theta \times \nonumber \\
\times exp \{-i [\omega_k(t_{\sigma + \tau/2}- t_{\sigma - \tau/2}) -k_x(x_{\sigma + \tau/2}- x_{\sigma - \tau/2})\;] \;\}.
\end{eqnarray}
Variables $t_{\sigma \pm \tau/2}$ and $x_{\sigma \pm \tau/2}$ are defined in (\ref{eq:trajectory}).\\
\indent Obviously, this expression depends on proper time $\sigma$, and transformation (\ref{eq:variables_change}) does not help to get rid of this  dependence because of the factor $\sin^2 \theta$ in its second integral. So observers which are at rest at different points of the detector in
 the hyperbolic trajectory observe completely different radiation powers, although there is no preferred time for such movement. So the case $\sigma=0$ considered in \cite{2021_Lynch} is a very special one and does not represent the properties of a uniformly accelerating detector. Nevertheless, it is useful to consider this case because all theoretical conclusions and experimental considerations done by LCHK are associated with this case.\\
\indent For $\sigma=0$ we have from \footnote{This expression corresponds exactly to $\Gamma$ in \cite{2021_Lynch} (S14) 
\begin{eqnarray}
\label{eq:DRP special case 0}
\Gamma= q^2 \frac{1}{(2 \pi)^3} \frac{1}{4}\int d\xi (1-\cosh a \xi)\int \frac{d^3k}{\omega}
\sin^2 \theta
e^{-i(\Delta E \xi +\omega \Delta t)}
\nonumber
\end{eqnarray}
 }
(\ref{eq:transverse DRP}) and (\ref{eq:trajectory})
\begin{eqnarray}
\label{eq:DRP special case}
S_{tr}(\sigma=0)= q^2 \frac{1}{4}
\int_{-\infty}^{+\infty}d\tau \exp(-i \Delta E \tau)\int \frac{d^3k}{(2 \pi)^3}\times
 \exp\{-i \frac{2}{a}\omega\sinh(a\tau)\}\times 
 \sin^2 \theta (1- \cosh a \tau )
\end{eqnarray}
\indent It can be reduced to  
\begin{eqnarray}
\label{eq:transverse power}
S_{tr}(\sigma=0)= \frac{2}{3}\alpha \frac{i}{\pi} \frac{a^2 }{2}\int_{0}^{+\infty} du 
\frac{u^{-i \Delta E/a}  (u-1)^2 u^{-1/2}.
	}{(1-u)^3}
\end{eqnarray}
the same way as it is done in \cite{2021_Lynch} (S14)-(S20)
This expression coincides with \cite{2021_Lynch}(S20) but despite this after integration we come to (Appendix \ref{evaluation})
\begin{eqnarray}
\label{eq:transverse power 1}
S_{tr}(\sigma=0)=\frac{1}{3}\alpha a^2 \frac{e^{2 \pi \Delta E/a}-1}
{e^{2 \pi \Delta E/a}+1} 
\end{eqnarray}
while LCHK, after integrating the same expression, obtained a different formula \cite{2021_Lynch}(S21)
\begin{eqnarray}
\label{eq:thermalized larmor}
S_{LCHK}(\eta=0)= \frac{2}{3}\alpha a^2 \frac{1}{1+e^{2 \pi \Delta E /a}}.
\end{eqnarray}
\indent The LCHK result is obviously wrong when considering a boundary condition. From (\ref{eq:transverse power}) we can see that its principle value integral is zero when $\Delta E=0$, and
$S_{tr}(\sigma=0, \Delta E =0) =0$. The expression (\ref{eq:transverse power 1}) satisfies this condition, but the latter one does not:
$S_{LFCHK}(\eta=0,\Delta E=0) \ne 0$. More details are provided in Appendix \ref{LCHK error}. \\
\indent Therefore, LCHK's statement, based in particular on  (\ref{eq:thermalized larmor}), that "we discover a Larmor formula and power spectrum that are both thermalized by acceleration" is not supported by their calculations.\\
\indent  We must add that  expression (\ref{eq:DRP special case 0}) and folowing from it (\ref{eq:transverse power}), which we have obtained in our approach, are obtained by LCHJK by "ad hoc", after some miscalculations. Following their approach you can not arrive at that result. This leaves hope to preserve their main claim about acceleration induced thermality by making the necessary adjustments in their approach to the problem. We discuss this issue in detail in Section \ref{The Thermal Larmor Formula}.\\
\subsection{LCHK's Derivation of Detector Radiation Expression}
\label{The Thermal Larmor Formula}
\indent LCHK start with the equation \cite{2021_Lynch} (S7) 
\begin{eqnarray}
\label{eq:P_equation}	
P= \int d^4 x d^4 x^{\prime}| <E_f|\hat{j}_{\mu}|E_i> |^2 <0|\hat{A}^{\dagger \mu}(x^{\prime})\hat{A}^{\nu}(x)|0>,
\end{eqnarray}
where integrals are taken over $(x)=(t,x,y,z)$ and $(x^{\prime})=(t^{\prime},x^{\prime},y^{\prime},z^{\prime})$ in the laboratory inertial reference frame, and immediately move on to the next equation
\begin{eqnarray}
\Gamma =\frac{d P}{d \eta}=q^2 \int d \xi e^{-i \Delta E \xi} U_{\mu \nu}(x^{\prime},x)G_{\mu \nu}(x^{\prime},x),
\end{eqnarray}
( $\xi=\tau^{\prime}-\tau$, $\eta=(\tau^{\prime}+\tau)/2$, and $\tau$, $\tau^{\prime}$ are proper times of the accelerating detector) \\
making some implicit assumptions. Discussion of these assumptions is the subject of this section. To expose this we need to go into some details of LCHK's calculations. \\
\indent LCHK use the matrix element \cite{2021_Lynch}(S5) 
\begin{eqnarray}
|<E_f|\hat{j_{\mu}}(x) |E_i>|^2= q^2 u_{\mu}(x^{\prime})u_{\nu}(x)\delta^3(x^{\prime}-x_{tr}^{\prime}(\tau^{\prime})) \delta^3(x-x_{tr}(\tau)) e^{-i \Delta E (\tau^{\prime}-\tau)}\nonumber \\
\vec{x}_{tr}(\tau)=(0,0,\frac{1}{a}\cosh(a \tau)), \; u_{\mu}=(\cosh(a \tau), 0,0, \sinh(a \tau)),
\end{eqnarray} 
which is a function of proper times $\tau$ and $\tau^{\prime}$,  in (\ref{eq:P_equation}) to integrate it over $\vec{x}$, $\vec{x}^{\prime}$, t and $t^{\prime}$ in the lab system. This creates some ambiguity. More uncertainty also comes from  semi-classical character of the model. The detector is endowed with two contradictory properties. It is both a quantum and a classical object. Its quantum properties are defined in a proper reference frame and presented by the factor $e^{-i \Delta E(\tau^{\prime}-\tau)}$.
Classical properties are defined by space-time coordinates $(t,x_1,x_2,x_3)$ in the lab system in which the quantum nature of the detector is ignored and the detector is considered as a point-like object. This approach is accepted in the literature, see for example \cite{1975_Davis}, and we employ it to further explore the capabilities of the proposed LCHK model. To support their idea we have to represent the matrix element in a different form
\begin{eqnarray}
|<E_f|\hat{j_{\mu}}(x) |E_i>|^2= q^2 u_{\mu}(x^{\prime})u_{\nu}(x)\delta^3(x^{\prime}-x_{tr}^{\prime}(t^{\prime})) \delta^3(x-x_{tr}(t)) e^{-i \Delta E (\tau^{\prime}-\tau)} \\
\label{eq:hyperbolic trajectory}
\vec{x}_{tr}(t)=(0,0,(1/a)(1+a^2t^2)^{1/2}),
\end{eqnarray} 
where all "classical" quantities related to the detector motion in 4-space-time must be considered as functions of $(t,\vec{x})$ rather than $\tau$. Then after integration over $\vec{x}$ and $\vec{x}^{\prime}$ we come to the equation \footnote{
	This step was skipped by LCHK. Therefore we use abbreviation, cor, for new or corrected expressions where our results are different from LCHK's ones. }
\begin{eqnarray}
\label{eq:LCHKnottations}
P_{cor}=\int d t^{\prime} dt\; q^2 u_{\mu}(t^{\prime}, 0,0,x_{tr,3}^{\prime}(t^{\prime})) u_{\nu}(t, 0,0,x_{tr,3}(t))e^{-i \Delta E (\tau^{\prime}-\tau)}<0|\hat{A}^{\dagger \mu}(t^{\prime}, 0,0,x_{tr,3}^{\prime}(t^{\prime})) \hat{A}^{\nu}(t, 0,0,x_{tr,3}(t))|0> \nonumber \\
\end{eqnarray} 
where all variables are defined on trajectory (\ref{eq:hyperbolic trajectory}) as functions of t. Particularly
\begin{eqnarray}
u_{\mu}=((1-v^2)^{-1/2},\vec{v}(1-v^2)^{-1/2} ), \;\; \vec{v}=(0,0, a t(1+a^2t^2)^{-1/2})
\end{eqnarray}
In this expression $\tau$ and $ \tau^{\prime}$ should be considered as parameters associated with quantum character of the detector and which are not connected with a specific classical trajectory of the detector. Our next step is to connect both classical and quantum features of the model and consider $(t,t^{\prime})$ as functions of $(\tau, \tau^{\prime})$. However, it is more convenient to work with proper time variables $(\eta,\xi)$:
\begin{eqnarray}
\label{eq:detector coordinates}
\tau^{\prime}= \eta +\xi/2, \;\;\tau= \eta -\xi/2,\nonumber  \\
t^{\prime}=(1/a)\sinh (a(\eta+\xi/2)), \;\;t=(1/a)\sinh (a(\eta-\xi/2)).
\end{eqnarray}
Then 
\begin{eqnarray}
\label{eq:detector coordinates 1}
x_{tr,3}^{\prime}(t^{\prime})=(1/a) \sinh(a(\eta+\xi/2)), \;\; x_{tr,3}(t)=(1/a) \sinh(a(\eta-\xi/2)), \nonumber   \\
u_{\mu}(t^{\prime}, 0,0,x_{tr,3}^{\prime}(t^{\prime})) =(\cosh(a(\eta +\xi/2)), 0,0, \sinh(a(\eta +\xi/2))) \nonumber \\
u_{\mu}(t, 0,0,x_{tr,3}(t)) =(\cosh(a(\eta -\xi/2)), 0,0, \sinh(a(\eta -\xi/2)))
\end{eqnarray}
The Jacobian of a transformation from $(t,t^\prime)$ to $(\eta-\xi/2,\eta+\xi/2)$ is  
\begin{eqnarray}
\label{eq:jacobian}
J(t, t^{\prime};\eta, \xi)=\cosh(a(\eta+\xi/2))\cosh(a(\eta-\xi/2))
\end{eqnarray}  
In the Coulomb gauge adopted in \cite{2021_Lynch}, (S11), the zero component of the operator $\hat{A}^{\mu}$ is ignored  \footnote{ This operation is in contradiction with LCHK statement  \cite{2021_Lynch}(S22) that their consideration is fully relativistic. }, and  equation (\ref{eq:LCHKnottations}), taking into account  (\ref{eq:jacobian}), becomes 
\begin{eqnarray}
P_{cor}=\int d\eta d\xi\; q^2 \;\cosh(a(\eta+\xi/2))\cosh(a(\eta-\xi/2)) \; \sinh(a(\eta+\xi/2))\sinh(a (\eta -\xi/2)) \times \nonumber \\
\times e^{-i \Delta E \xi}
<0|\hat{A}^{\dagger 3}(t^{\prime}, 0,0,x_{tr,3}^{\prime}(\eta+\xi/2)) \hat{A}^3 (t, 0,0,x_{tr,3}(\eta-\xi/2))|0>
\end{eqnarray}
or using the expansion of $\hat{A}^3 (t, 0,0,x_3(\tau))$ (\ref{eq:expansion}) in plane waves and properties of polarization vectors (\ref{eq:polarization vectors})
\begin{eqnarray}
P_{cor}=\int d\eta d\xi \;
q^2 \cosh(a(\eta+\xi/2))\cosh(a(\eta-\xi/2)) \sinh(a(\eta+\xi/2))\sinh(a(\eta-\xi/2)) \;
e^{-i \Delta E\; \xi}\times \nonumber \\
\times \int\frac{d^3k}{(2 \pi)^3} \frac{1}{2 \omega_k} e^{-i \omega_k(t^{\prime}-t)+ i k_3( x_{tr,3}^{\prime}(t^{\prime})-x_{tr,3}(t) )}\times \sin^2\theta.
\end{eqnarray}
where variables \; t, $t^{\prime}$, $x_{tr,3}^{\prime}(t^{\prime})$, and $x_{tr,3}(t)$
are defined in (\ref{eq:detector coordinates}) and {\ref{eq:detector coordinates 1}) as functions of $(\eta+\xi/2)$ and $(\eta-\xi/2)$, and $\theta$ is an angle between wave vector $\vec{k}$ and the direction of the detector motion along axis $x_3$. \\
\indent This corrected expression is  different from the corresponding expression, \cite{2021_Lynch} (S13), by an additional factor, Jacobian
	$J(t, t^{\prime};\eta, \xi)$.
	Because this step was skipped by LCHK, they lost this factor. Then 
	\begin{eqnarray}
	\label{eq:integral}
	\Gamma_{cor} =\frac{d P_{cor}}{d \eta}=q^2 \frac{1}{(2 \pi)^3}\frac{1}{16}\int d\xi\; [2 \cosh^2(2 a \eta)-1-\cosh(2 a\xi)] \times \nonumber \\
	\times
	\int\frac{d^3k}{ \omega_k} 
	\sin^2 \theta
	e^{-i \Delta E\;\xi-i \omega_k(t^{\prime}-t)+ i k_3(x_3^{\prime}(t^{\prime})-x_3(t))}.
	\end{eqnarray}
	Following LCHK, we set $\eta=0$, even though this assumption contradicts to the feature of the hyperbolic motion that it has no preferred time. Then 
	\begin{eqnarray}
	\Gamma_{cor}(\eta=0)=q^2 \frac{1}{(2 \pi)^3}\frac{1}{16}\int d\xi\; [1-\cosh(2 a\xi)] \times 
	\int\frac{d^3k}{ \omega_k} 
	\sin^2 \theta e^{-i \Delta E\;\xi-i \omega_k(t^{\prime}-t)} = \nonumber \\
	= \frac{1}{4} (\frac{2}{3} \alpha\frac{1}{2 \pi}) \int d\xi (1 - \cosh2 a \xi) e^{-i(\delta E)\xi} \int d \omega \omega
	e^{-i \frac{2 \omega}{a} \sinh(a \xi /2)}, \;\;\; \alpha \equiv\frac{q^2}{4 \pi}
	\end{eqnarray}
	and finally, weighting the frequency integral with an additional factor of frequency, 
	\begin{eqnarray}
	\label{eq:LCHKcorrected 0}
	S_{cor}= \frac{1}{4} (\frac{2}{3} \alpha\frac{1}{2 \pi}) \int d\xi (1 - \cosh2 a \xi) e^{-i(\Delta E)\xi} \int d \omega \omega^2
	e^{-i \frac{2 \omega}{a} \sinh(a \xi /2)}, \;\;\; \alpha \equiv\frac{q^2}{4 \pi}
	\end{eqnarray} 
	This result is significantly different from \cite{2021_Lynch}(S16)
	\begin{eqnarray}
	\label{eq:radiated power}
	S_{LCHK}=\frac{2}{3}\alpha \frac{1}{2 \pi}\int d\xi e^{-i\Delta E \xi}(1-\cosh a\xi)\int d \omega\; \omega^2 
	e^{-i \omega \Delta t}, \;\;
	\Delta t=\frac{2}{a} \sinh(a \xi/2)
	\end{eqnarray}
	It has $\cosh(2 a \xi)$ instead of $\cosh(a \xi)$ and an additional factor, $\frac{1}{4}$.\\ 
	\indent Having integrated (\ref{eq:LCHKcorrected 0}) we come to 
	\begin{eqnarray}
	\label{eq:LCHKcorrected}
	S_{cor}=S_{cor,1} + S_{cor,2}=
	(\frac{2}{3}\alpha a^2)\;\frac{1}{4} \frac{e^{2\pi\Delta E/a}-1}{e^{2 \pi \Delta E/a}+1}-
	(\frac{2}{3}\alpha a^2) \frac{i}{4 \pi} \int_{0}^{+\infty}dw \frac{w^{-1/2 - i \Delta E /a}}{(w-1)^2}
	\end{eqnarray}
	This expression has two parts. The first one, convergent $S_{cor,1}$, is significantly different from the so the called "thermalized Larmor formula"
	\begin{eqnarray}
	\label{eq:thermolized larmor}
	S_{LCHK}(\Delta E)=\frac{2}{3} \alpha a^2 \times\frac{1}{1 + e^{2 \pi \Delta E/a}}
	\end{eqnarray}
	The second one, $S_{cor,2}$, is divergent 
	and it is absent in LCHK's calculations. This demonstrates that even analysis carried out as closely as possible to the approach accepted by LCHK leads us to a totally different result from their "thermalized Larmor formula" \cite{2021_Lynch}(S21).
	
\section{Discussion}
 The transition between theoretical predictions and experimental observations made by LCHK is based on their model of the Unruh-DeWitt detector, coupled to a semi-classical 4-vector current. We would expect this model to satisfy at least two conditions. It should predict and/or explain possible results of experiments and it  must correctly represent the theoretical concept to be proved.  In the foregoing work we have considered the LCHK model only from the theoretical point of view. \\
\indent Our analysis was focused on two issues: first, whether LCHK's claim that they "discover[ed] a Larmor formula and power spectrum that are both thermalized by the acceleration" is supported by the calculations and second, whether this model is consistent and has no internal contradictions in the frame of quantum electrodynamics.\\
\indent For this purpose we examined possible expressions of a detector radiation power, its derivations and interpretations, and associated  two-potential correlation functions for a quantized electromagnetic field observed at the location of an accelerating detector. We have considered two cases in the frame of the Lorentz gauge, with and without 
use of the Lorentz complementary condition, and analyzed in great detail LCHK's approach made in the Coulomb gauge. \\
\indent Radiation power expression (\ref{eq:detector radiation power }), corresponding to correlation function (\ref{eq:correlation function integral }), is easy to interpret. It is the same for any observer at a location of the detector on a hyperbolic trajectory, an expected feature of hyperbolic motion. But, as our calculations show, it includes non-physical states and is divergent. So, (\ref{eq:detector radiation power }) is not a good representation of the thermal effects phenomenon.\\
\indent Expression (\ref{eq:transverse DRP}) for DRP, obtained in the Lorentz gauge frame and under the complementary condition, includes only physical, transverse, polarization modes. However, it depends on detector proper time. This contradicts the fact that there is no preferred time for hyperbolic motion. Therefore its calculation in a general form does not make sense. \\
\indent Nevertheless, calculation of (\ref{eq:transverse DRP}) for a very special case,  
 when proper time and detector lab velocity is zero, can easily be done and leads us to
(\ref{eq:transverse power}). This expression was also considered by LCHK, though as we 
have indicated in Subsection \ref{The Thermal Larmor Formula} was obtained ad hoc, and after its integration our result  
 (\ref{eq:transverse power 1}) is totally different from the mistaken LCHK result (\ref{eq:thermalized larmor}). Radiation power (\ref{eq:transverse power 1}) shows some signs of thermality which could be associated with a detector acceleration 
but they are different from the Bose-Einstein statistics expected for the photon field. We have shown that if the detector energy gap is zero, $\Delta E=0$, then there is no radiation,
$S_{tr}(\sigma=0)$, and, in complete contradiction of the LCHK claim, no "thermalized Larmor formula".  \\
\indent There remains one more feature of the model to be discussed. To separate transverse modes from scalar/longitudinal ones we use a gauge transformation. This can only be done in one certain reference frame and 
it is impossible to do it in all three reference frames $I_M$,$I_{\tau_1}$
 and $I_{\tau_2}$,
 involved in the calculations of the correlation functions, simultaneously. For example, after a gauge transformation  (\ref{eq:gauge transformation}),  4-potentials (\ref{eq:after gauge transformation})  have only transverse polarization modes. They are defined in the inertial lab system $I_M$. However, these potentials being considered in $I_{\tau_1}$ and
$I_{\tau_2}$ by means of Lorentz transformations from $I_M$
to $I_{\tau_1}$ and $I_{\tau_2}$ with different boosts, correspondingly $\gamma_{\tau_1}$ and $\gamma_{\tau_2}$, would have contained both transverse and scalar/longitudinal polarization modes. Thus the theory of this detector cannot be constructed based on transverse polarizations only, and non-physical modes are also involved. \\
\indent The only very special case (\ref{eq:DRP special case}) for the LCHK model with a finite valued result (\ref{eq:transverse power 1}) was obtained. It has the following features: 1. there is no certain thermality, 2. the DRP expression is not Lorentz-invariant and represents only one point of the classical hyperbolic trajectory,
3. it involves both transverse and non physical longitudinal/scalar modes, in contradiction with the features of a quantized electromagnetic field,  4. a detector, representing a quantum device, moves along a classical trajectory. This is in contradiction with the quantum uncertainty principle. \\
 \indent These features dictate how well the model answers DeWitt's question mentioned in the introduction: "How would a given particle detector respond in a given situation?" 
 Based on our analysis, we must conclude that LCHK's model cannot provide a theoretical justification of a quantum electrodynamic phenomenon being discussed.
\appendix
\section{Vector Potentaials, Polarization vectors and Hyperbolic Functions}
\label{Vector Potentials}
\indent  In the reference frame $I_M$, the expansion of the operator of a 4-potential of the field in plain waves is \cite{1994_Milonni},(10.166):
\begin{eqnarray}
\label{eq:expansion}
\hat{A}_{\mu}(t,\vec{x})=\int \frac{d^3k}{(2 \pi)^3}\frac{1}{2 \omega_k}\sum_{\lambda=0}^{3}[a(\vec{k},\lambda)e^{-kx} +a^{+}(\vec{k},\lambda)e^{kx}] e_{\mu}(\vec{k},\lambda), 
\end{eqnarray}
where annihilation/creation operators obey the commutation relationship
\begin{eqnarray}
\label{eq:cr_an_operators}
[a(\vec{k,\lambda}),a^{+}(\vec{k}^{\prime},\lambda^{\prime}))]=-(2 \pi)^3 2 \omega_k g^{\lambda \lambda^{\prime}}\delta(\vec{k}-\vec{k}^{\prime}), \;\;\;
g^{\lambda \lambda^{\prime}}=(+1,-1,-1,-1).
\end{eqnarray}
and polarization vectors have the following properties
\begin{eqnarray}
\label{eq:polarization vectors}
\sum_{\lambda=0}^{3}e_\mu(\vec{k},\lambda)e_\nu(\vec{k},\lambda) g^{\lambda \lambda}=-g_{\mu \nu}, \;\; \{g_{\mu \nu}\}=\{g_{\lambda \lambda^{\prime}}\}=(1,-1,-1,-1),\nonumber \\
e_{\mu}(\vec{k}, \lambda=0)=(1,0,0,0),\;\; e_{\mu}(\vec{k}, \lambda=1)=(0,\vec{e}(\vec{k},\lambda=1)), \nonumber \\
e_{\mu}(\vec{k}, \lambda=2)=(0,\vec{e}(\vec{k},\lambda=2)), \;\; e_{\mu}(\vec{k}, \lambda=3)=(0,\vec{k}/k), 
\nonumber \\
\vec{e}(\vec{k},\lambda) \vec{e}(\vec{k},\lambda^{\prime})
=\delta_{\lambda \lambda^{\prime}},\;\; \lambda, \lambda^{\prime}=1,2,3
\nonumber \\
\vec{e}(\vec{k},1)\times\vec{e}(\vec{k},2)=\vec{e}(\vec{k},3),\;\;\vec{e}(\vec{k},2)\times\vec{e}(\vec{k},3)=\vec{e}(\vec{k},1),\;
\;\vec{e}(\vec{k},3)\times\vec{e}(\vec{k},1)=\vec{e}(\vec{k},2)
\end{eqnarray}
Here are two useful formulas for hyperbolic functions used in this paper:
\begin{eqnarray}
\label{eq:hyperbolic functions}
\cosh[a(\sigma+\frac{1}{2}\tau)]\times \cosh[a(\sigma-\frac{1}{2}\tau)]= \frac{1}{2} \cosh(2a\sigma)+\frac{1}{2} \cosh(a\tau), \nonumber \\
\sinh[a(\sigma+\frac{1}{2}\tau)]\times \sinh[a(\sigma-\frac{1}{2}\tau)]= \frac{1}{2} \cosh(2a\sigma)-\frac{1}{2} \cosh(a\tau)
\end{eqnarray}
\section{Error in LCHK' calculation of integral (\ref{eq:transverse power})}
\label{LCHK error}
Integral in \cite{2021_Lynch}(S20)

\begin{eqnarray}
\label{eq:I_LCHK}
S_{LCHK} = \frac{2}{3}\alpha \frac{i}{\pi} (\frac{a}{2})^3 \frac{8}{a}
\int_{0}^{\infty} dw \frac{w^{1/2 -i\Delta E/a - \frac{1}{2}w^{3/2-i \Delta E/a }- \frac{1}{2}w^{-1/2 -i \Delta E/a} 	}}{(w-1)^3}
\end{eqnarray}
at $\Delta E=0$ becomes 

\begin{eqnarray}
I(\Delta E =0) = (-1/2) \int_{0}^{\infty} dw \frac{w^{-1/2}}{w-1}.
\end{eqnarray}
After variable change $w=1/u$ we get the same expression with an opposite sign
\begin{eqnarray}
I(\Delta E =0) = (+1/2) \int_{0}^{\infty} du \frac{u^{-1/2}}{u-1}.
\end{eqnarray}
So $I(\Delta E =0)=0$, and $S_{LCHK}$ defined in (\ref{eq:I_LCHK}) should be 0 at $\Delta E=0$. But their final formula \cite{2021_Lynch}(S21)}
\begin{eqnarray}
S_{LCHK}=(2/3)\alpha a^2 \frac{1}{1+e^{2 \pi \Delta E/a}}
\end{eqnarray}
does not satisfy this boundary condition. 

 This result was used in Section (\ref{transverse DRP}) 

\section{Evaluation of the Integral in (\ref{eq:transverse power}).}
\label{evaluation}

\begin{wrapfigure}{l}{0.5\textwidth}
	\centering
	\includegraphics[scale=0.4]{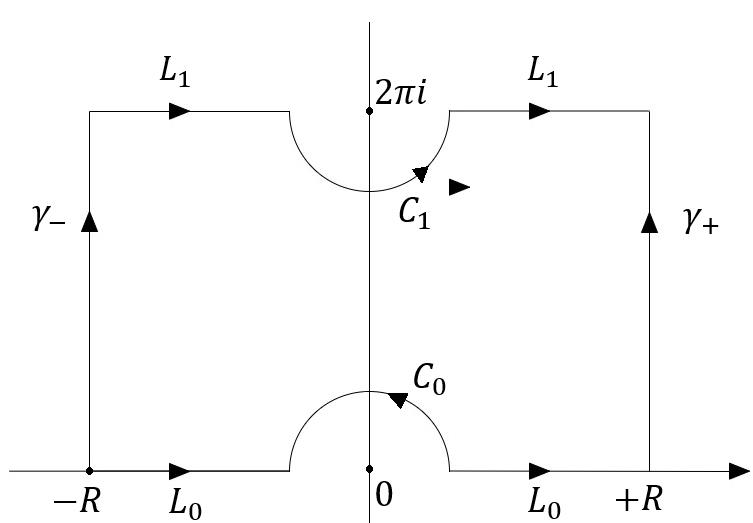}
		\caption{Integration contour}
\end{wrapfigure} 

The principle value integral in (\ref{eq:transverse power})  
\begin{eqnarray}
\label{eq:integralI1}
I \equiv \int_{0}^{+ \infty} \frac{u^{-i \frac{\Delta E}{a}-\frac{1}{2}}}{1-u}du 
\end{eqnarray}
after a variable change  has a form
\begin{eqnarray}
I=\int_{-\infty}^{+\infty} \frac{e^{t(1/2-ib)}dt}{1-e^t},\;\; u=e^t, \;\; b=\frac{\Delta E}{a} 
\end{eqnarray}
For integration purposes we use a complex variable function 
$f(z)=\frac{e^{z(1/2-ib)}}{1-e^z}$ and integration contour $\gamma=L_0- L_1 + \gamma_{+} - \gamma_{-} -C_0 -C_1 $ 
\\ \\ \\
 where $R\rightarrow +\infty$
in Fig 1. By the residue theorem we have
\begin{eqnarray}
0=\int_{\gamma}\; f(z) dz= (\int_{L_0} - \int_{L_1} + \int_{\gamma_{+}} -\int_{\gamma_{-}}- \int_{C_0} -\int_{C_1}) f(z)dz, \nonumber
\end{eqnarray} 
 where 
\begin{eqnarray}
\int_{L_0} f(z)dz \rightarrow I, \;\;\; \int_{L_1} f(z)dz \rightarrow
 \int_{-\infty}^{+\infty}\frac{e^{(t+2 \pi i)(1/2-ib)} dt}{1-e^t} =-Ie^{2 \pi b}, \;\;
\int_{\gamma_{+}} f(z)dz\rightarrow 0, \int_{\gamma_{+}} f(z)dz\rightarrow 0,  \nonumber \\
\int_{C_0}f(z)dz \rightarrow i \pi Res(f(z), 0)= -i \pi , \;\;\; \int_{C_0}f(z) \rightarrow i \pi Res(f(z), 0)= i \pi \; e^{2 \pi b } \nonumber
\end{eqnarray}
Then we obtain
\begin{eqnarray}
I=\frac{i \pi (e^{2 \pi \Delta E/a }-1)
}{e^{2 \pi \Delta E/a }+1}
\end{eqnarray}
\section{Divergence of $I_2$ defined in (\ref{divergent integral})}
\label{Divergence of I2}
Integral $I_2$ in (\ref{divergent integral}) has a singular point at $\tau=0$. Its principle value, if it exists, should be defined as a sum of two integrals  
\begin{eqnarray} 
I_2 =\lim_{\epsilon \rightarrow 0}\int_{-\infty}^{-\epsilon}d\tau \frac{-i \sin( \Delta E \tau) \cosh(a\tau)}{\sinh^3(\frac{a \tau}{2})}
 + \lim_{\epsilon \rightarrow 0}\int_{+\epsilon}^{+\infty}d\tau \frac{-i \sin( \Delta E \tau) \cosh(a\tau)}{\sinh^3(\frac{a \tau}{2})} \equiv I_{21}+I_{22}
\end{eqnarray}
But none of these limits exists. More exactly, both limits are positive infinite, and the integral is divergent.\\
\indent For example, integral $I_{21}$ would have existed if for any small positive $\delta$ there exists such positive $\eta$ that
\begin{eqnarray}
| \int_{-\epsilon^{\prime}}^{-\epsilon{\prime \prime}}d\tau \frac{-i \sin( \Delta E \tau) \cosh(a\tau)}{\sinh^3(\frac{a \tau}{2})}|< \delta, \;\;\; \epsilon^{\prime \prime}< \epsilon^{\prime}
< \eta
\end{eqnarray}
But it is easy to see that it is impossible. The integral tends to infinity when $\epsilon^{\prime}$ and $\epsilon^{\prime \prime}$ become small. Indeed,
\begin{eqnarray}
| \int_{-\epsilon^{\prime}}^{-\epsilon{\prime \prime}}d\tau \frac{-i \sin( \Delta E \tau) \cosh(a\tau)}{\sinh^3(\frac{a \tau}{2})}| = 
| \int_{-\epsilon^{\prime}}^{-\epsilon{\prime \prime}}d\tau \{ 
\frac{2\sin( \Delta E \tau) }{\sinh(\frac{a \tau}{2})}+
\frac{\sin( \Delta E \tau) }{\sinh^3(\frac{a \tau}{2})}
\}|= \nonumber \\
| \frac{2\Delta E  }{a/2}(-\epsilon^{\prime \prime}+ \epsilon^{\prime}) +
\frac{\Delta E}{(a/2)^3}(\frac{1}{\epsilon^{\prime \prime}}-\frac{1}{\epsilon^{\prime}})| 
\rightarrow +\infty 
\end{eqnarray}
\section{Acknolegments}
I would like to thank prof. A. Tumanov for helpful discussion of some mathematical issues in this paper.  


\section{Bibliography}
\begin{thebibliography} {99}
\bibitem{1975_Davis}Davis, P.C.W., Scalar Particle Production in Schwarzschild and Rindler Metrics, J. Phys.A8, 609 (1975)
\bibitem{1976_Unruh} W.G. Unruh, "Notes on Black-Hole Evaporation", Phys. Rev.D14, 870 (1976)
\bibitem{1982_Birrell} Birrell, N.D.,Davis, P.C.W. Quantum Fields in Curved Space.(1982)  
\bibitem{1994_Milonni} Milonni, P.W. The Quantum Vacuum. An Introduction to Quantum Electrodynamics ( 1994 )
\bibitem{1979_DeWitt} De-Witt. Quantum Gravity: The new synthesis, pages 680-745, in the book "General Relativity. An Einstein Century Survey".
( 1979 ) by S.W. Hawking, W. Israel  
\bibitem{2017_Cozzella} Cozzella, G., Landulfo,A. G. S., Matsas,G. E. A., Vanzella,D. A. T. Proposal for observing the Unruh effect using classical electrodynamics, Phys. Rev. Lett. 118, 161102 (2017).
\bibitem{2021_Lynch} Lynch, M.H., Cohen, E., Hadad, Y.,Kaminer,I. Experimental Observation of Acceleration Induced Thermality
Phys. Rev. D 104, 025015 ( 2021 ) 
\bibitem{2022_Lynch} Lynch,M.H. Notes on the Experimental Observation of the Unruh Effect. https://inspirehep.net/literature/2081864
\bibitem{1964_Dirac}Dirac, P.A.M.The Principles of Quantum Mechanics.
\bibitem{1980_Boyer}Boyer,T.H. Phys.Rev. D21, No 8, 2137 (1980), Phys.Rev. D29, No 6, 1089 (1984)
\bibitem{1985_Cole} Cole,D.C. Phys. Rev. D31, No 8, 1972 (1985), Phys. Rev. D35, No 2(1987) 
\bibitem{1962_Jackson}Jackson, J.D.Classical Electrodynamics  (1962)
\bibitem{2013_Mihailov} Mihailov,V.D. Theory of Coimplex Variable Functions. Praktikum. Ministry of Education and Science of the Russion Federation.Nuclear Research Institute, 2013 (In Russian.)
\bibitem{1973_Landau} Landay, L. D. and Lifshits, Y. M. (1973). The Field Theory.
\bibitem{1961_Schweber} Schweber,S.S. An Introduction to Relativistic Quantum Field Theory (1961)
\end{thebibliography}
\end{document}